\documentclass[twocolumn,letterpaper,amsmath,amssymb]{revtex4}
\usepackage{graphicx}
\usepackage{dcolumn}
\usepackage{bm}
\begin{document}

\def\vri{\vec{r}_{i}}
\def\vrj{\vec{r}_{j}}
\def\rij{r_{ij}}
\def\vrij{\vec{r}_{ij}}
\def\drij{\hat{r}_{ij}}
\def\vdr{\delta\vec{r}}
\def\dr{\delta{r}}
\def\s{\hat{s}}

\title{Small world-Fractal Transition in Complex Networks: Renormalization Group Approach}

\author{Hern\'an D. Rozenfeld, Chaoming Song and Hern\'an A. Makse}

\affiliation {Levich Institute and Physics Department, City College of
New York, New York, NY 10031, US}

\begin{abstract} 

 {\bf 
We show that renormalization group (RG) theory applied to complex networks are useful to classify network topologies into universality classes in the space of configurations. The RG flow readily identifies a small-world/fractal transition by finding (i) a trivial stable fixed point of a complete graph, (ii) a non-trivial point of a pure fractal topology that is stable or unstable according to the amount of long-range links in the network, and (iii) another stable point of a fractal with short-cuts that exists exactly at the small-world/fractal transition. As a collateral,  the RG technique explains the coexistence of the seemingly contradicting fractal and small-world phases and allows to extract information on the distribution of short-cuts in real-world networks, a problem of importance for information flow in the system. 
}
\end{abstract}
\maketitle

A generic property that is usually inherent in scale-free networks but applies equally
well to other types of networks, such as in Erd\H{o}s-R\'enyi random graphs, is the \emph{small-world}
feature~\cite{erdos,albert99}.
In small-world networks a very small number of steps is required to reach a given
node starting from any other node. This is expressed by the slow (logarithmic) increase
of the average diameter of the network, $\bar{r}$, with the
total number of nodes $N_0$, $\bar{r} \sim \ln N_0$, where $r$
is the shortest distance between two nodes through network links. 

The small-world property has been shown to apply in many empirical studies of diverse
systems.
However, recent work~\cite{song05,goh06,song06,jskim07} showed that many networks that have been found to display the small-world property, such as the WWW, are indeed fractal, indicating a power-law dependence of the distances with the network size, $\bar{r} \sim N_0^{1/d_B}$, where $d_B$ is the fractal dimension, up to a certain length-scale before the global small-world behavior is observed.
Therefore, it is not clear how it is possible that fractal scale-free networks coexist with the small world property.
This shows the need for a mathematical framework that reconciles these two seemingly contradictory aspects, fractality and the small-world property.

In this paper we show
that renormalization group (RG) theory, previously developed to
understand critical phase transitions in physical systems
\cite{kadanoff} and recently extended to inhomogeneous networks~\cite{song05,goh06,song06,jskim07}, provides such a framework. The main result of our work is four-fold: (1) We introduce a method based on the RG technique that classifies network topologies into three universality classes according to fixed points of the RG flow. We find a stable trivial fixed point of a complete graph, a non-trivial fixed point of a fractal structure that becomes stable or unstable according to the amount of long-range links added to the network, and a third stable fixed point that exists exactly at the small-world/fractal transition consisting of a fractal with short-cuts.
(2) The RG technique allows for finding the distribution of short-cuts overlaying a pure fractal topology, a technique that we test in real-world networks like the WWW and biological networks. (3) The RG identifies a second point which is associated with information flow in the system.
(4) The RG analysis finds an explanation for the seemingly contradiction between the small-world effect observed at a global scale in real-world networks and the fractal behavior occurring at finite scales.

\begin{figure}[ht]
\centering {
\resizebox{0.48\textwidth}{!}{\includegraphics{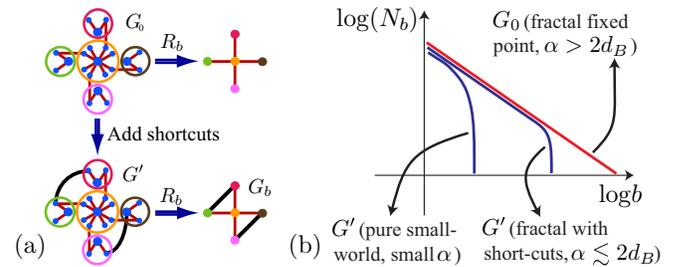}}
}
\caption{(a) Top panel: fractal network $G_0$ with box covering for $b=2$, and its renormalized network $R_b(G_0)$. Bottom panel: short-cuts are added (black links in lower left panel) we obtain network $G'$. We apply the RG transformation to $G'$ and obtain $G_b$ (lower right panel).
(b) Sketch of the number of boxes versus the box diameter according to the probability, $P(r) = Ar^{-\alpha}$, to add short-cuts between nodes at distance $r$. When the network is in the stable phase, $s\equiv \alpha/d_B >2$, RG flows toward the pure fractal  fixed point $G_0$, where the number of boxes is a power-law for all values of $b$ (red straight line). When $s \lesssim 2$, the number of boxes displays a power-law with an exponential cut-off at large values, indicating that globally the network topology behaves as a small-world, but for small values of $b$ the network still exhibits fractality. For small values of $\alpha$, the number of boxes decays exponentially and the network displays the small-world property at all scales.
}
\label{net}
\end{figure}

We apply the RG to complex networks using the box-covering technique~\cite{song05}. The network is covered with boxes such that all nodes within a box are at a distance smaller than $b$ (top panel of Fig.~\ref{net}a), where distance is
 the number of links along the shortest path between two nodes. Once the network is tiled, we construct the renormalized network: by replacing each box with supernodes (or renormalized nodes) and these supernodes are connected if there is at least one link between two nodes in their corresponding boxes. When this RG transformation, $R_b$, is applied to a network $G_0$, it leads to a new network $G_b$. If $G_0$ is self-similar, as it was empirically shown to be the case for the WWW and many biological networks~\cite{song05,goh06,song06,jskim07}, the RG leads to a structure that presents similar properties as $G_0$. More technically, if $G_0$ is a fractal network, then $R_b(G_0)=G_0$ and $G_0$ is a fixed point of the RG flow.

Suppose we start with the fractal network $G_0$ and add short-cuts according to the distance $r$ between nodes with probability $p(r) = A r^{-\alpha}$, where $r>1$. The new network with short-cuts, $G'$, is not self-similar anymore or in other words, $R_b(G') \neq G'$ (see Fig.~\ref{net}a for a simple example of this process). Here we show that depending on the value of the exponent of the short-cuts $\alpha$, the application of the RG process brings $G'$ either back to the original self-similar structure $G_0$ or transforms it into a complete graph (where all nodes are connected to each other). $G_0$ and the complete graph are both \emph{fixed points} in the space of networks with a substantial difference between them. $G_0$ is an \emph{unstable} fixed point of $R_b$ since a small perturbation (small number of short-cuts) may lead it to a drastically different network under $R_b$. The complete graph is a \emph{stable} or trivial fixed point because any small perturbation always returns the network into the complete graph under $R_b$.

We start by calculating the RG flow in the space of configurations.
Let $d_B$ be the box (or fractal) dimension of the original self-similar network $G_0$. Thus $b^{d_B}=N_0/N_b$ is the average number of nodes in a box of
size $b$ where $N_b$ is the number of nodes in $G_b$ (or number of boxes in $G_0$) and $N_0$ the number of nodes is $G_0$.
After a renormalization step is applied to the network $G'$, the
probability to find a short-cut between two nodes at distance $r$ in
the renormalized network  $G_b$ (black links in lower right panel of Fig.~\ref{net}a) is $p_b(r) = 1 -
(1-p(br))^{b^{2d_B}}$, and therefore,
\begin{equation}
p_b(r) = 1 - \left[1-A (br)^{-\alpha} \right]^{b^{2d_B}}.
\end{equation}
For simplicity we write $x \equiv A^{-1} (br)^{\alpha}$ and
$B(r)\equiv A^{2d_B/\alpha}~r^{-2d_B}$. After repeatedly applying the
renormalization transformation, i.e. $b \to \infty$, we find a fixed
point of the RG flow defined at
\begin{eqnarray}
p^*(r) &\equiv& \lim_{b\rightarrow\infty} p_b(r) = 1 -
\left[\lim_{x\rightarrow\infty}
  \left(1-\frac{1}{x}\right)^{B(r)~x^{2d_B/\alpha}}\right] \nonumber \\ 
  &=& 1 -
\lim_{x\rightarrow\infty}
  \exp\Big[-B(r)~x^{2d_B/\alpha-1}\Big].
  \label{fixed}
\end{eqnarray}

\begin{figure}[th]
\centering {
\resizebox{0.48\textwidth}{!}{\includegraphics{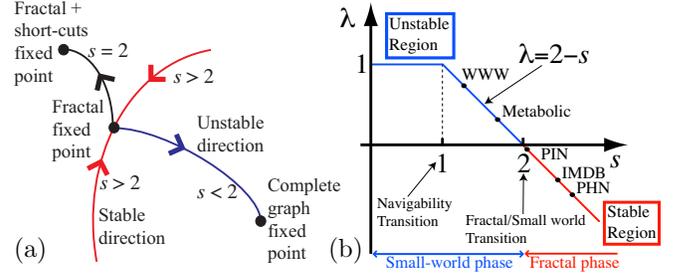}}
}
\caption{
(a) RG flow diagram. 
(b) Phase diagram of the three
phases found via RG at $s < 1$, $1<s <2$, and $s > 2$.
}
\label{diagram}
\end{figure}

Analysis of Eq.~(\ref{fixed}) reveals the existence of a critical
value at $s \equiv \alpha/d_B = 2$ separating two
phases of the RG flow.
If $s > 2$, we find $p^*(r) = 0$. Therefore, the RG flow
brings the network toward the self-similar fixed point $G_0$, implying that the
added short-cuts disappear under the renormalization flow. 
On the other hand, if $s < 2$, we find $p^*(r) = 1$.  In this
case, $G'$ flows under RG toward a
trivial fixed point consisting of a complete graph where all the nodes are
connected to each other.  If $s = 2$, $G'$
flows toward another non-trivial
stable fixed point consisting of the original fractal network $G_0$ with short-cuts following $p^*(r) = 1 - \exp(-A r^{-2d_B})$ (Fig~\ref{diagram}a).

To better understand the features of the phases identified by the RG
flow, we analyze the behavior of the average network degree under
renormalization. This calculation allows one to identify a second
critical point within the unstable phase related to information flow in the system.
Let $z_0$ be the average degree (number of
links per node) of the unperturbed network, $G_0$, and $z'$ the
average degree of $G'$ after the short-cuts are added. Then, $z' -
z_0 = \frac{2M(L)}{N_0}$, where $M(L)$ is the number of short-cuts at
distance $L$ (the diameter of $G_0$). Since $G_0$ is fractal, we
find,
\begin{equation}
M(L) \approx d_B \int_{1}^{L} A r^{-\alpha} r^{d_B -1}{\rm d}r = 
\frac{A}{1 - s} (L^{d_B(1-s)}-1).
\end{equation}
Hence we obtain
\begin{equation}
z' - z_0 = 
\frac{2A}{1 - s} \Biggl(\frac{L^{d_B(1-s)}-1)}{N_0} \Biggr).
\end{equation}

After renormalizing the network $G'$ with length-scale $b$, short-cuts
connecting nodes inside a box will not appear in the renormalized
network, $G_b$. Therefore, the number of remaining short-cuts in $G_b$
is simply the number of short-cuts that connect different boxes, i.e.
$M(L)-M(b)$.  If $z_b$ is the average degree of the renormalized
network, $G_b$, then
\begin{equation}
z_b - z_0  =  \frac{2(M(L)-M(b))}{N_b} =  (z' - z_0) f_N(b),
\end{equation}
where
\begin{equation}
f_N(b)= \Biggl(\frac{L^{d_B(1-s)}-b^{d_B(1-s)}}{L^{d_B(1-s)}-1}\Biggr) b^{d_B}.
\end{equation}
We define the renormalization parameter $x_b \equiv N_0 / N_b = b^{d_B}$, and in the limit of large networks, $L \rightarrow \infty$,  we find the scaling:
\begin{equation}
\label{av_deg}
f_N(x_b)  \sim x_b^{\lambda},
\end{equation}
where the RG exponent $\lambda$ depends on the long-range exponent
$\alpha$ as
\begin{eqnarray}
\label{av_deg_lambda}
\lambda = \left\{ \begin{array}{ll}
  1, & {\rm if}~~s \leq 1,\\
   2 - s, & {\rm if}~~s > 1.
\end{array} \right.
\end{eqnarray}

Equation (\ref{av_deg_lambda}) (see Fig.~\ref{diagram}b) identifies two transitions separating
different phases in the space of configurations, as depicted in the
phase diagram of Fig.~\ref{diagram}. The first transition at
$s=2$ corresponds to the point when $\lambda = 0$, and
separates a stable phase with $\lambda <0$ for $s > 2$
from an unstable phase with $\lambda > 0$ for $s <
2$. Therefore this transition corresponds to the complete graph/fractal transition
identified by the analysis of Eq.~(\ref{fixed}),
and corresponds to the point
at which the network topology dramatically changes. In the unstable
phase, $s < 2$, the average degree increases ($\lambda > 0$), so that
under infinite steps of the RG procedure the network becomes a
complete graph with infinite average degree in the thermodynamic
limit. On the other hand, when $s > 2$, the the network conserves the
global fractal structure of $G_0$. Under the RG flow the difference between $z_b$ and $z_0$ goes to 0 and the short-cuts
disappear, returning $G'$ back to its original fractal structure. 
In this state, the diameter of the
network grows as a power law with the number of nodes, implying a
large-world fractal structure. 
Instead, if long-range connections are added with $s < 2$, the
small-world property is achieved, where the diameter of the network
grows logarithmically with the number of nodes. Therefore, the
$s=2$ (or $\alpha = 2d_B$) transition is a small-world/fractal transition.  This calculation generalizes the small/large-world transition, previously found in Refs.~\cite{benjamini01,coppersmith02,kozma05} for lattices to the case of complex networks.

An important point readily emerges from the analysis of
Eq.~(\ref{av_deg_lambda}) when $s = 1$ ($\alpha = d_B$)
within the unstable phase, as shown in Fig.~\ref{diagram}a. 
Notice that $s = 1$ coincides with the optimal point of decentralized navigability of Kleinberg~\cite{kleinberg00}
for lattices with fractal dimension $d_B$~\cite{roberson06}, and therefore this results could be seen an a plausible extension of the results in Refs.~\cite{kleinberg00,roberson06} for scale-free complex networks. We expand on this point later.

As a test of the RG predictions we use a model of fractal networks, as described in
Ref.~\cite{song06}.
Using the fractal model, short-cuts with an exponent $\alpha$ can be added to the network and the prediction of Eq.~(\ref{av_deg_lambda}) can be tested in a controlled manner.

The fractal model network is built as follows~\cite{song06}: At
generation $n=0$, we start with a star network of 5 nodes, i.e. a node in the center and four nodes connected to the center node.
Then,
generation $n+1$ is obtained recursively by attaching $m$ new nodes to
the endpoints of each link $l$ of generation $n$. 
In addition, 
we remove links $l$ of generation $n$ and add $x$ new links connecting
pairs of new nodes attached to the endpoints of $l$ (see top panel of
Fig.~\ref{net}a for an example at with $n=2$, $m=2$, and $x=1$). The
algorithm leads to a pure fractal scale-free network with degree
distribution exponent $\gamma=1+ {\rm ln}(2m+x)/{\rm ln}~2$ and
fractal dimension $d_B = {\rm ln}(2m+x)/{\rm ln}~3$.

\begin{figure}[th]
\centering {
\resizebox{0.48\textwidth}{!}{\includegraphics{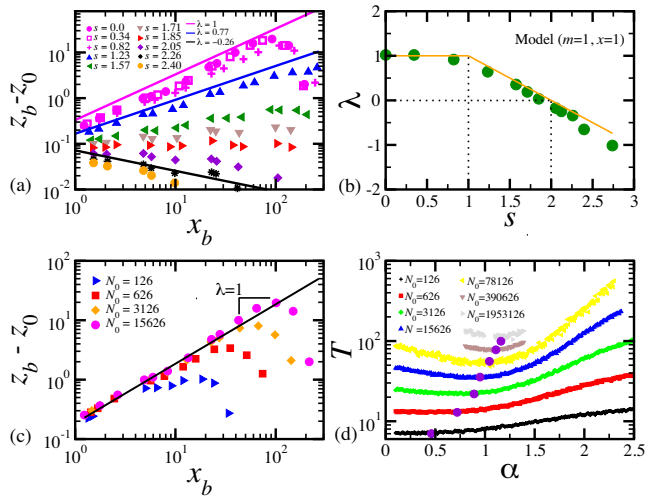}}
}
\caption{Renormalization applied to the
 fractal model. (a) Average degree of the renormalized network
 versus $x_b$ for the fractal model with $n=6,m=2,x=1$.
 (b) Values of $\lambda$ obtained from Fig.~\ref{renorm}a (green data
 points) and the prediction of the RG from Eq.~(\ref{av_deg}) (solid
 line). 
 (c) Finite size effects in $z_b - z_0$.
 (d) Navigation time $T(\alpha)$ for the fractal network model.
 Circles show the minimum value. 
 }
\label{renorm}
\end{figure}

Figure~\ref{renorm}a shows the results of the RG flow applied to the
fractal model network with $n=6,m=2,x=1$ with $d_B=1.46$ for
various long range exponents $\alpha$.  Starting from a pure fractal
topology we pick a node and add a random connection to another node at
distance $r$ according to the probability $p(r) = A r^{-\alpha}$
when $r>1$ (in our results we repeat this process for 10\% of the nodes in the network, 
although our conclusions are independent of this value.)
The renormalization is performed numerically
using the box covering algorithm called MEMB in Ref.~\cite{song07}. Notice that
the MEMB algorithm leads to networks that are smaller than the original, 
and therefore few points are obtained in the plot of $z_b - z_0$ vs. $x_b$. 
To overcome this problem and obtain better resolution, we take advantage of 
the self-similar aspect of the network and perform a ``partial renormalization'', 
in which parts of the network are subsequently renormalized into supernodes.
We follow the behavior of $z_b$ in the RG flow for a given $\alpha$.  For
$s<1$, the average degree follows a power-law with
exponent $\lambda=1$ as in Eq.~(\ref{av_deg_lambda}). When $s > 1$ the exponent follows the
theoretical prediction Eq.~(\ref{av_deg_lambda}),
$\lambda=2-s$. Figure~\ref{renorm}b shows a very close comparison with
theory indicating the transition between the stable region and the
unstable region and the optimal navigability point.
In Fig.~\ref{renorm}c we show the dependence of $z_b - z_0$ for different values
of $N_0$. When $b$ is large, finite size effects become evident and the average degree of the
renormalized network deviates from the expected power-law of Eq.~(\ref{av_deg}).

The fractal model also allows us to verify directly that the point
$s = 1$ could be regarded as a plausible extension of the optimal point of decentralized navigability 
found by Kleinberg in lattices~\cite{kleinberg00}.
We test this by measuring numerically the average time, $T(\alpha)$, 
for a message to be delivered from a source node to a target
node along the links of the network. 
It is important to notice that since scale-free networks
are not embedded in any euclidean space, one cannot directly apply the 
decentralized algorithm as introduced by Kleinberg. In the
case of scale-free networks, we allow for the message holder to have information 
on the distance between \emph{any} node and the target in the fractal background $G_0$, 
but not on their long-range short-cuts that exists in $G'$.
In Fig.~\ref{renorm}c we show simulation results for
$T(\alpha)$ versus $\alpha$ for the fractal model with $m=2,x=1$
and for different values of system size $N_0$. We find
that the value $\alpha_c(N_0)$ corresponding to the minimum delivery
time for a given $N_0$ slowly converges to the critical value
$\alpha_c(N_0) \to \alpha_c = d_B = 1.46$ as $N_0 \to \infty$, 
implying a navigability transition at $s=1$ as predicted by the RG analysis. 
A finite size scaling analysis of $\alpha_c(N_0)$ versus the
inverse square of the logarithm of $N_0$, as suggested in Ref.~\cite{roberson06}, confirms this result.

An advantage of the RG approach is that it allows for a
measurement of the type of short-cuts present in real-world networks. 
Previously in this paper, we started with a pure fractal structure $G_0$ to which
short-cuts were added, generating network $G'$, and analyze the stability
of $G'$ under the RG procedure. We now switch to the study of real-world
networks where we tackle the inverse problem. The real-world networks
we examine are known to have an underlying fractal structure since the
measurement of $N_b$ versus $b$ leads to a power-law relation~\cite{song05}
(see Fig.~\ref{net}b).
However, these real-world networks already present short-cuts overlaying
the fractal structure, and therefore rather than a pure power-law, 
the scaling shows a cut-off at large $b$, like $G'$ in Fig.~\ref{net}b.
Therefore, these networks are 
composed by a fractal underlying structure, analogous to $G_0$,
with some short-cuts generating the network $G'$. 
The question we want to answer here is, what is 
the $\alpha$ exponent of the short-cuts overlaying the fractal network?
Since one cannot obtain the value of $\alpha$ directly from the data 
(as it is not possible to distinguish a priori between the links of the 
fractal structure and the short-cuts links) we infer its value by treating the real-world 
network as the network $G'$ and measuring directly the 
value of $\lambda$ using the RG flow.

\begin{figure}[t]
\centering {
\resizebox{0.45\textwidth}{!}{\includegraphics{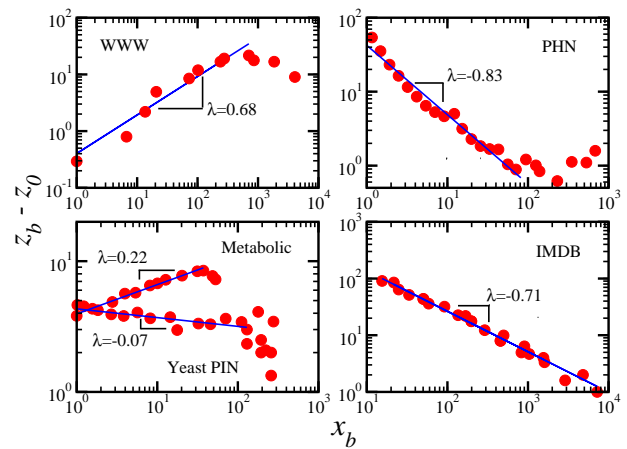}}
}
\caption{
Average degree of the renormalized
 network versus $x_b \equiv N_0 / N_b = b^{d_B}$ 
 for the studied samples.
 }
\label{realnets}
\end{figure}
In Fig.~\ref{realnets} we show the results of the RG flow to a sample of the WWW~\cite{albert99}, a protein
homology network~\cite{arnold05}, the metabolic network of {\it E. coli}~\cite{almaas04}, a yeast
protein interaction network~\cite{mering}, and the co-acting network of IMDB that have been found to exhibit fractal
topologies~\cite{song05}. 
Table~\ref{table1} shows a summary of the exponents for real-world networks.
For instance we find that while the WWW exhibits fractal scaling in $N_b$~\cite{song05} it also
presents enough short-cuts that its structure belongs to the unstable phase. 
Thus, the WWW is fractal up to a given length scale and then it crosses over to small world behavior
at large scales ($N_b$ presents an exponential cut-off at large $b$, Fig.~\ref{net}b).
In fact, for large values of $b$ we observe a deviation from the power-law, which may be attributed to 
the finite size of these networks,
similar to what we observe in Fig.~\ref{renorm}c for the fractal model at different sizes.
The RG determines the crossover scale such that under enough RG steps the WWW finally becomes a complete graph.
\begin{table}
\caption{Exponents obtained for the real-world networks.}
\label{table1} 
\begin{tabular}{|c|c|c|c|c|c|c|}
\hline Network & $d_B$ from~\cite{song05} & $\lambda$ (Fig.~\ref{realnets}) & $s$ from Eq. (\ref{av_deg_lambda}) & phase\\ 
\hline 
WWW & 4.1  &  0.68  & 1.32 & unstable \\ 
\hline 
Metabolic & 3.5 & 0.22 & 1.78  & unstable \\
\hline 
PIN & 2.2  & -0.07  & 2.07 & stable \\ 
\hline 
IMDB & 3.88 & -0.71 & 2.71  & stable \\ 
\hline 
PHN & 2.5  & -0.83  & 2.83 & stable \\ \hline \noalign{\smallskip}
\end{tabular}
\end{table}
Thus, the renormalization allows to conceptualize the apparent discordance between small-world and fractal properties. 
Also, due to the its proximity to the $\alpha = d_B$ point, the WWW is sufficiently randomized to give a topology close to optimal information flow.
On the contrary, the biological networks PHN and PIN, and the social network of co-acting belong to the stable phase
indicating that the short-cuts are minimal (the metabolic network is unstable but close to the transition point).
The biological networks display a modular deterministic structure shaped by evolution which exhibit pure fractal character that may be seen as a means of protection, preservation and conservation.

In summary, the RG approach finds the type of short-cuts in a
given network and determines the location of a network in the space of configurations. 
When the exponent of the short-cuts is $\alpha > 2d_B$, the network structure belongs to the stable
phase, where RG exhibits a fixed point consisting of a pure fractal network in the space of configurations. 
On the other hand, when $\alpha < 2d_B$, the network is in the unstable 
phase where short-cuts become dominant, changing dramatically the 
global distance between nodes, and leading to a small-world network at large scales. 

{\it Acknowledgments.} We thank L. Gallos and C. Briscoe
for fruitful discussions. This work was supported by NSF grants SES-0624116 and EF-0827508.


\begin{thebibliography}{99}

\bibitem{erdos}
P. Erd\H{o}s and A. R\'enyi,
{\it Publ. Math. Inst. Hung. Acad. Sci.} {\bf 5}, 17 (1960).
D. Watts and S. Strogatz,
{\it Nature} {\bf 393}, 440 (1998).



\bibitem{albert99} R. Albert, H. Jeong, and A. -L. Barab\'asi,
{\it Nature} {\bf 401}, 130 (1999).

\bibitem{song05} C. Song, S. Havlin, and H. A. Makse,
{\it Nature} {\bf 433}, 392 (2005).

\bibitem{goh06} K. I. Goh, G. Salvi, B. Kahng, and D. Kim,
{\it Phys. Rev. Lett.} {\bf 96}, 018701 (2006).

\bibitem{song06} C. Song, S. Havlin, and H. A. Makse,
{\it Nature Phys.} {\bf 2}, 275 (2006).

\bibitem{jskim07} J. S. Kim {\it et al.},
{\it Phys. Rev. E} {\bf 75}, 016110 (2007).
F. Radicchi, J. J. Ramasco, A. Barrat, and S. Fortunato,
{\it Phys. Rev. Lett.} {\bf 101}, 148701 (2008).



\bibitem{kadanoff} L. P. Kadanoff, 
{\it Statistical Physics: Static, Dynamics and Renormalization} 
(World Scientific, Singapore, 2000).

\bibitem{benjamini01} I. Benjamini and N. Berger,
{\it Random Structures \& Algorithms} {\bf 19}, 102 (2001).

\bibitem{coppersmith02} D. Coppersmith, D. Gamarnik, and M. Sviridenko,
{\it Rand. Struct. and Alg.} {\bf 21}, 1 (2002).

\bibitem{kozma05} B. Kozma, M. Hastings, and G. Korniss,
{\it Phys. Rev. Lett} {\bf 95}, 018701 (2005).

\bibitem{kleinberg00} J. Kleinberg,
{\it Nature} {\bf 406}, 845 (2000).

\bibitem{roberson06} M. R. Roberson and D. ben-Avraham,
{\it Phys. Rev. E} {\bf 74}, 017101 (2006).

















\bibitem{song07} C. Song, L. K. Gallos, S. Havlin, and H. A. Makse, 
{\it J. Stat. Mech.: Theory and Experiment} {\bf 03}, 03006 (2007).



\bibitem{arnold05} R. Arnold {\it et al.},
{\it Bioinformatics} {\bf 2}, 42 (2005).

\bibitem{almaas04} E. Almaas, B. Kovacs, T. Vicsek, Z. N. Oltvai, and A. -L. Barabasi,
{\it Nature} {\bf 427}, 839 (2004).

\bibitem{mering} C. von Mering {\it et al.},
{\it Nature} {\bf 417}, 399 (2002).











\end{thebibliography}
\end{document}